\documentclass{emulateapj}
\usepackage{apjfonts,graphics}

\newcommand{\simgt}{\lower.5ex\hbox{$\; \buildrel > \over \sim \;$}}
\newcommand{\simlt}{\lower.5ex\hbox{$\; \buildrel < \over \sim \;$}}

\newcommand{\baredth}{\;\overline{\raise1.0pt\hbox{$'$}\hskip-6pt \partial}\;}
\newcommand{\edth}{\;\raise1.0pt\hbox{$'$}\hskip-6pt\partial\;}

\newcommand{\lan}{\langle}
\newcommand{\ran}{\rangle}

\newcommand{\btheta}{\mbox{\boldmath $\theta$}}

\lefthead{Umetsu et al.}
\righthead{Cluster Baryon Fraction and Structure}

\begin{document}

\title{Cluster Baryon Fraction and Structure from the 
Convergence/SZ Effect Diagram}
 
\author{Keiichi Umetsu\altaffilmark{1}, 
Jun-Mein Wu\altaffilmark{2},
Tzihong Chiueh\altaffilmark{2},
Mark Birkinshaw\altaffilmark{3}}
\altaffiltext{1}{Institute of Astronomy and Astrophysics, Academia
Sinica,  P.~O. Box 23-141, Taipei 106,  Taiwan, Republic of China}
\altaffiltext{2}{Department of Physics, National Taiwan University, Wenshan Chiu, Taipei 116, Taiwan}
\altaffiltext{3}{Department of Physics, University of Bristol, Tyndall Avenue, Bristol BS8 1TL, UK.}

\begin{abstract}
The cross-correlation of Sunyaev-Zel'dovich effect (SZ)
and weak-lensing imaging surveys can be used to test how 
well hot baryons trace dark matter in clusters of galaxies.
We examine this concept using mock
SZ and weak-lensing surveys based on the forthcoming 
AMiBA experiment and
generated from a pre-heated cosmological $N$-body/hydrodynamic
simulation. A cross-correlation diagram between matched lensing
convergence and Compton $y$ maps exhibits butterfly-wing-like
structures, corresponding to individual clusters, that
encode rich
information about the distributions of hot gas and dark matter. When
the cluster redshift and temperature are available the slope of a wing
reveals the cluster gas fraction and the width of the wing indicates
how badly the hot gas traces dark matter. On the basis of simulated data we
discuss systematic errors in the projected gas fraction estimates that
would be obtained from such survey comparisons. 
\end{abstract}


\section{Introduction}

Clusters of galaxies, the largest virialized systems known,
   are key tracers of the matter distribution in the Universe.
   Clusters are composed of the hot, diffuse intracluster medium (ICM),
   luminous galaxies (stars), and a dominant mass component in the
   form of the unseen dark matter. The mass ratio of hot gas to dark
   matter in clusters (the gas fraction) is believed to be a fair
   estimate of the universal baryon fraction (White et
   al. 1993). However, non-gravitational processes associated with cluster
   formation, such as radiative gas cooling or pre-heating, would
   break the self-similarities in cluster properties and hence cause 
   the gas fraction to acquire some mass dependence (Kravtsov et
   al. 2005; Bialek et al. 2001).
Weak gravitational lensing on background galaxies 
probes directly the projected cluster mass distribution 
(Bartelmann \& Schneider 2001).
The thermal 
Sunyaev-Zel'dovich (SZ) effect on the cosmic microwave background 
spectrum, on the other hand, 
measures the projected
thermal electron pressure (Birkinshaw 1999).
Therefore, 
a combination of weak-lensing and SZ measurements will allow us to
measure directly the cluster gas fractions 
when the ICM
temperature is available
(Myers et al. 1997; Holder, Carlstrom, \& Evrard 2000).

Both blind SZ surveys, such as that to be done with AMiBA (Ho et
   al. 2004), and gravitational weak lensing surveys, such as the
   CFHT Legacy Survey\footnote{http://www.cfht.hawaii.edu/Science/CFHTLS}, aim to detect massive clusters of
   $\simgt 10^{14} M_{\odot}$ to study the formation of structure
   through the abundance and properties of the detected clusters.
   A comparison of the two types of survey should reveal much 
   about the different structures of hot gas and dark matter in
   clusters of different mass. 
Here we investigate this possibility using $N$-body simulations.

\section{Cluster gas fractions from weak-lensing and SZ observations}


The lensing convergence $\kappa$ is essentially the surface mass
density projected on the sky,
\begin{equation}
\label{eq:kappa}
\kappa
= \Sigma_m/\Sigma_{\rm crit}(z_d,z_s),
\end{equation}
where 
$\Sigma_m$ is the surface mass density of a halo acting as
gravitational lens of redshift $z_d$, and 
$\Sigma_{\rm crit}$
is  the lensing critical surface mass density 
as a function of 
lens redshift 
$z_d$ and source redshift $z_s$
specified for a given background cosmology
(e.g., Bartelmann \& Schneider 2001).

The SZ effect is described by the 
Compton $y$ parameter defined as
a  line-of-sight integral of the temperature-weighted 
thermal electron  density (Birkinshaw 1999):
\begin{equation}
\label{eq:y}
y=
 \frac{\sigma_T}{m_e c^2}\int\!dl\, n_e k_{\rm B}T_g \approx
 \frac{1+X}{2}
\frac{\sigma_T}{m_p} \frac{k_{\rm B}T_g}{m_e c^2}\Sigma_g.
\end{equation}
where  $n_e$ is the electron number density, $T_g$ the gas
temperature, $m_e$ the electron mass, $\sigma_{\rm T}$ the Thomson cross
section, $\Sigma_g$ the surface gas mass density,
and $X=0.76$ the hydrogen abundance.
Note that we have assumed isothermality in the second
equality in Eq.~(\ref{eq:y}).
The presence of temperature gradients in the ICM
still remains controversial  (see Piffaretti et al. 2005 and references
therein).
In the paper, 
we assume the ICM is isothermal as a zeroth-order approximation; 
we found 
a temperature variation within the virial radius 
by a factor two at most
from our sample of massive halos (see \S 4).





In general,
the lensing $\kappa$ and the Compton
$y$ reconstructed from noisy observations are smoothed by a 
certain window function
in the map making process.
In interferometric observations, the synthesized beam (or PSF) is well
approximated by a Gaussian window. In weak lensing surveys, Gaussian
smoothing is often adopted to find clusters as local maxima in the
reconstructed convergence map (e.g., Miyazaki et al. 2002). 
We hence express the convergence field 
smoothed with a Gaussian window $W_{\rm G}(\theta)= \exp(-\theta^2/\theta_{\rm G}^2)/\pi\theta^2_{\rm
G}$ as $\kappa_{\rm
G}(\btheta)$;
$\theta_{\rm G}$ is related to FWHM of the Gaussian window
$\theta_{\rm fwhm}$ by
$\theta_{\rm G}=\theta_{\rm fwhm}/\sqrt{4\ln(2)}$.
Similarly, we define
the Gaussian smoothed Compton $y$ parameter, $y_{\rm G}(\btheta)$.
We then compare the reconstructed $\kappa_{\rm G}$ and
$y_{\rm G}$ values at a certain point $\btheta$.
Taking the ratio of smoothed $\kappa_{\rm G}$ and $y_{\rm G}$,
we have
\begin{eqnarray} 
\label{eq:y2k}
\eta_g \equiv 
\frac{y_{\rm G}}{\kappa_{\rm G}}
&=&
8.9h\times 10^{-4}
\left( \frac{k_{\rm B}T_g}{10{\rm keV}} \right)
\left( \frac{\Sigma_{\rm crit}}{1h\, {\rm g/cm^2}} \right)
\left(\frac{f_g}{0.1}\right)
\end{eqnarray}
where $f_g(\btheta)  = \Sigma_{g,{\rm G}}(\btheta)
/\Sigma_{m,{\rm G}}(\btheta) $ is the {\it local} gas fraction
defined with the Gaussian smoothed surface mass densities.
%
In order to extract information on the cluster gas fraction from
weak lensing and SZ observations, we thus need additional 
information on the ICM temperature 
$T_g$, the cluster redshift $z_d$,
and the redshift distribution of background galaxies.
Such information can be
in principle
available from
X-ray observations and photometric redshift measurements of background
galaxies.

We can define a {\it global} estimator for the cluster gas fraction 
based on weak-lensing and SZ observation as
\begin{equation}
\label{eq:eta}
\lan \eta_g \ran \equiv\frac
  {\sum_i w_i \eta_{g,i}}
  {\sum_i w_i}\propto \Sigma_{\rm crit}(z_d,z_s)\, \lan f_g T_g \ran
\end{equation}
where $\eta_{g,i}=\eta_g(\btheta_i)$ and $w_i=w(\btheta_i)$
is the weight for the $i$th pixel $\btheta_i$.
Accordingly, the global gas fraction of a halo can be estimated
from $\kappa$ and $y$ maps
as $\lan f_g \ran \propto \lan \eta_g \ran T_g^{-1}\Sigma_{\rm crit}^{-1}$
if one
assumes 
the isothermality of the ICM.
The case with $w_i=\kappa_{\rm G}(\btheta_i)$ corresponds to a commonly used definition
of the gas fraction, $\eta_g^{(1)}\equiv
\lan y\ran /\lan \kappa\ran 
\propto
\lan \Sigma_g\ran /\lan \Sigma_m\ran $, as the ratio of the mean
surface mass densities. However, the mean convergence $\lan \kappa\ran $
is ill constrained from weak-lensing observations:
Weak lensing $\kappa$ reconstructions
based solely on image distortions
suffer from the {mass sheet} degeneracy 
(Schneider \& Seitz 1995)
unless additional information
such as the magnification bias is taken into account
(Broadhurst, Taylor \& Peacock 1995; Broadhurst et al. 2005).
On the other hand, interferometric observations are insensitive to
the DC signal, and hence do not constrain the total flux in the
field-of-view (FoV). Bolometers also suffer from
a high level of contamination by
atmospheric and other environmental signals, and hence rely on the
differencing scheme to extract the sky signals.
We therefore consider an alternative
 weight  $w_i=\kappa_{\rm G}^2(\btheta_i)$
that weights high density regions more strongly than $w_i=\kappa_{\rm G}(\btheta_i)$.
Then we have the gas fraction estimator of the form:
\begin{equation}
\label{eq:cor}
\eta_g^{(2)}  \equiv \frac
  {\sum_i \kappa_{\rm G}(\btheta_i)\,y_{\rm G}(\btheta_i)}
  {\sum_i \kappa_{\rm G}(\btheta_i)\,\kappa_{\rm G}(\btheta_i)} = 
\frac{\lan \kappa y\ran }{\lan \kappa^2\ran}.
\end{equation}
%
Suppose that $\kappa$ and $y$ satisfy a linear bias relation
   $y(\btheta)=a\,\kappa(\btheta)$. Then errors in the Compton
   parameter of the form $y\to y+\epsilon$ (where $y \gg |\epsilon|$)
   lead to errors in the two estimators of $\eta_g$ of
   $\Delta \eta_g^{(1)}=\epsilon/\lan \kappa\ran $ and 
   $\Delta \eta_g^{(2)}=\epsilon\lan \kappa \ran /\lan \kappa^2\ran $.
   The difference between these errors
   $|\Delta \eta_g^{(1)} |-|\Delta \eta_g^{(2)} |=
   \left(|\epsilon| / \lan\kappa \ran \right)
   \left({\rm Var}[\kappa]/ \lan \kappa^2\ran \right)\ge 0$, with
   equality only if variance in $\kappa$ vanishes. Thus the 
   $\eta_g^{(2)}$ estimator is always better than the 
   $\eta_g^{(1)}$ estimator. Similarly, an error on the convergence of
   the form $\kappa \to \kappa+\epsilon$ (where $\kappa \gg |\epsilon|)$
   leads to 
   $|\Delta \eta^{(1)}|-|\Delta \eta^{(2)}  |
   \approx 
   \left(a|\epsilon| / \lan\kappa \ran \right)
   \left({\rm Var}[\kappa]/ \lan \kappa^2\ran \right)\ge 0$, again
   showing that the cross-correlation based estimator $\eta_g^{(2)}$
   is more robust than $\eta_g^{(1)}$.


In practical observations, noise contributions to the estimator
   (\ref{eq:eta}) must be taken into account. Since the noise
   properties between the $\kappa$ and $y$ maps will not be
   correlated, the required correction is in the
   denominator of Eq.~(\ref{eq:cor}), where we replace
   $\lan \kappa^2 \ran$ by $\lan \kappa^2\ran -\sigma_{\kappa}^2$ with 
   $\sigma_{\kappa}^2$ being the noise variance in the smoothed
   convergence field, $\kappa_{\rm G}$ (\S 3.3).

\section{Mock observations}

We use cosmological simulation data by Lin et al. (2004)
to demonstrate what useful information 
we can extract from 
the cross correlation between weak-lensing and SZ surveys.
Specifically,
we simulate the forthcoming AMiBA experiment as an illustration
of such combined surveys.

\subsection{$N$-body/hydrodynamic simulations}\label{sect:data}

To make sky maps with realistic SZ and weak lensing signals,
we use results from preheating cosmological simulations
of a $\Lambda$CDM model 
$(\Omega_m=0.34,\Omega_{\Lambda}=0.66,\Omega_b=0.044,h=0.66,\sigma_8=0.94)$
in a $100h^{-1}$Mpc co-moving box 
generated with an N-body/hydrodynamics code
GADGET (Springel et al. 2001),
which reproduce the observed cluster $M_X$-$T_X$
and $L_X$-$T_X$ relations at $z=0$ (Lin et al. 2004).
We construct a total of $29$ SZ 
sky maps, each $1 {\rm deg}^2$ in size and containing $1024^2$ pixels,
by projecting the electron
pressure through the randomly displaced and oriented simulation boxes,
separated by $100h^{-1}$ Mpc, along a viewing cone out  to the redshift
of $z=2$. Similarly, 
29 $\kappa$-maps are constructed by projecting
the distance-weighted mass overdensity $\delta\rho$
out to a source plane 
at a fixed redshift, $z = z_s$.

\subsection{AMiBA SZ cluster survey}

As an AMiBA specification, we adopt
a close-packed hexagonal configuration of $19\times 1.2$m dishes
on a single platform.
AMiBA operates at $95$ GHz, and
this array configuration yields a synthesized beam of 
 $\theta_{\rm fwhm}\approx 2'$ ($0.6h^{-1}$Mpc at $z=0.8$),
which is optimized to detect high-$z$ clusters (Zhang et al. 2002).
The primary beam has an FoV of 
${\rm FWHM} \approx 11'$.
We assume the following telescope system: bandwidth $\Delta\nu=20$ GHz,
system temperature
$T_{\rm sys}=70$ K, system efficiency of $0.7$, and dual polarizations.

AMiBA will operate as a drift scan interferometer to remove
   ground contamination to high order (Pen et al.
   2002\footnote{http://www.cita.utoronto.ca/\~{}pen/download/Amiba};
   Zhang et al. 2002; Park et al. 2003).
   We approximate the resulting synthesized beam in the AMiBA map 
   by a Gaussian with $\theta_{\rm fwhm}=2'$, and we convolve 
   the simulated Compton $y$ maps with this Gaussian beam to generate a
   set of {\it noise-free} AMiBA $y_{\rm G}$-maps. The expected rms
   noise over a sky area of $\Omega_s$ with a total integration time
   of $t_{\rm int}$ is (Pearson et al. 2003),
   \begin{equation}
    \label{eq:ynoise}
    \sigma_{y_{\rm G}} =  
    2.0\times 10^{-6}
    \left(  \frac{T_{\rm sys}}{70{\rm K}} \right)\!
    \left(  \frac{\Delta\nu}{20{\rm GHz}} \right)^{-1/2}\!\!
    \left( \frac{\Omega_{\rm s}}{1{\rm deg}^2} \right)^{1/2}\!\!
    \left(   \frac{ t_{\rm int}}{ 240{\rm hrs}}\right)^{-1/2}.
   \end{equation}


\subsection{Weak lensing cluster survey}

We assume a weak lensing survey with a mean galaxy number
density of $n_g=40$ arcmin$^{-2}$,
an rms amplitude of the intrinsic
ellipticity distribution of $\sigma_{\epsilon}=0.3$,
and a mean galaxy redshift of $z_s=1$,
which are close to typical values for a ground based optical imaging
survey with a magnitude limit of $R_{\rm lim}\simgt 25.5$ mag in a
sub-arcsecond seeing condition (e.g., Hamana, Takada, \& Yoshida 2004).
The rms noise in a Gaussian-smoothed convergence field, $\kappa_{\rm
G}$,
is given as
\begin{equation}
\label{eq:knoise}
\sigma_{\kappa_{\rm G}}=1.1\times 10^{-2}
\left(\frac{\sigma_{\epsilon}}{0.3}\right)
\left(\frac{n_g}{40{\rm arcmin}^{-2}}\right)^{-1/2}
\left(\frac{\theta_{\rm fwhm}}{2'}\right)^{-1}
\end{equation}
(e.g., van Waerbeke 2000).
We choose the same smoothing scale of  $\theta_{\rm fwhm}=2'$ as 
for the AMiBA SZ survey and
convolve the $\kappa$ maps from $N$-body simulations
(see \S 3.1) with the Gaussian beam
to produce a set of {\it noise-free} $\kappa_{\rm G}$ maps.
Note that the adopted value of $\theta_{\rm fwhm}$ is
close to the optimal smoothing scale of 
$\theta_{\rm fwhm} \approx 1\farcm 7$ (or $\theta_{\rm G}\approx 1'$)
for an efficient survey
of massive halos with $M\simgt 10^{14}h^{-1}M_{\odot}$ found 
by Hamana et al. (2004) based on cosmological $N$-body simulations.

\begin{figure}
\epsscale{1.0}
\plotone{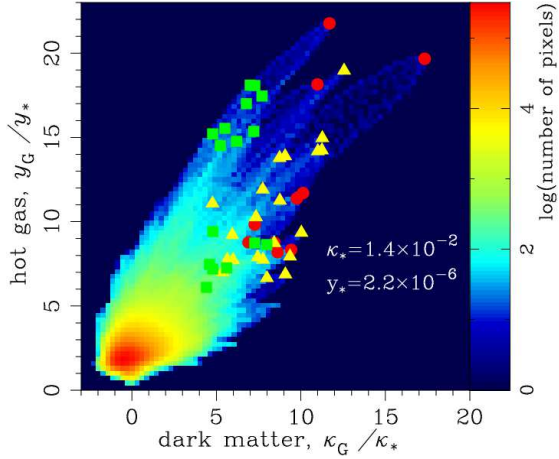}
\label{fig:buttfly_mock}
\caption{
Cross correlation between 
$\kappa_{\rm G}$ and $y_{\rm G}$ maps as a 2D scatter plot histogram
obtained from 
$29\times 1{\rm deg}^2$
{\it noise-free} sky maps of weak-lensing and SZ surveys for the AMiBA experiment.
The $\kappa_{\rm G}$ and $y_{\rm G}$ values are normalized in units
of rms dispersions  $\kappa_*$ and $y_*$, respectively.
Halo peaks with $\kappa \simgt 4.5\kappa_*$ and $y \simgt 6y_*$ are 
indicated with markers
according to their redshifts: 
$z<0.3$ ({\it circle}), $0.3\le z<0.5$ ({\it triangle}), $0.5\le z<1$
 ({\it square}).
}
\label{fig:bfd1}
\end{figure}

\section{Cluster gas fractions from 
cross correlation between weak-lensing and SZ signals}


We carried out a statistical analysis on the simulated 
weak lensing $\kappa_{\rm G}$ and SZ Compton $y_{\rm G}$ maps of 
$29\times 1{\rm deg}^2$ for the AMiBA experiment.
Figure 1 shows the cross correlation between 
the noise-free  $\kappa_{\rm G}$ and $y_{\rm G}$ sky maps
of the entire $29{\rm deg}^2$
as a 2D scatter plot histogram. The scatter plot is presented in units
of rms dispersions $\kappa_*$ and $y_*$ of the smoothed noise-free
fields $\kappa_{\rm G}(\btheta)$ and $y_{\rm G}(\btheta)$, respectively;
$\kappa_*=1.4\times 10^{-2}$ and $y_*=2.2\times 10^{-6}$.
Halo peaks with
$\kappa_{\rm G} \simgt 4.5\kappa_*$ and $y_{\rm G}\simgt 6y_*$
($\approx 3.3{\rm mJy/beam}$ at $95$ GHz) 
are indicated in Fig.~1 
according to their redshifts.
These thresholds effectively select massive halos
    with $M_{200} \simgt 2\times 10^{14}h^{-1}M_{\odot}$ (see Fig.~2 of
    Umetsu et al. 2004); $M_{\Delta_c}\equiv
    (4\pi/3)\Delta_c\rho_{c}r_{\Delta_c}^3$ is the mass enclosed by
    $r_{\Delta_c}$ with the critical density of the universe
    $\rho_c$.
Figure 1 clearly exhibits different capabilities of 
SZ and weak-lensing in cluster detections: weak lensing depends 
strongly on $z_d$ through $\Sigma_{\rm crit}$,
while the surface brightness of
SZ effect is independent of $z_d$.
In particular, gravitational lensing is less efficient
for high-$z$ lensing halos due to pure geometrical effects. 
As a consequence, high-$z$ halos
tend to appear in low-$\kappa$ and high-$y$ regions.

Another important feature in Fig.~1 is a number of
   butterfly-wing-shaped structures extending from the origin to
   points corresponding to the halo peaks. Each blade of a butterfly
   wing corresponds to an individual cluster. 
   Hereafter, we call
   the $\kappa/y$ cross-correlation diagram the {\it butterfly
   diagram}. For illustrative purposes, we show in Fig.~2 a butterfly
   diagram from the noise-free result for the most massive halo, which
   has $M_{200}=7.0\times 10^{14}h^{-1}M_{\odot}$ at $z=0.142$.
The SZ signal for this halo
is more extended than the weak lensing signal in this cluster
region. Furthermore, we can see a clear disparity between the 
shapes of the contours in the $\kappa$ and $y$ maps.

\begin{figure}
\plotone{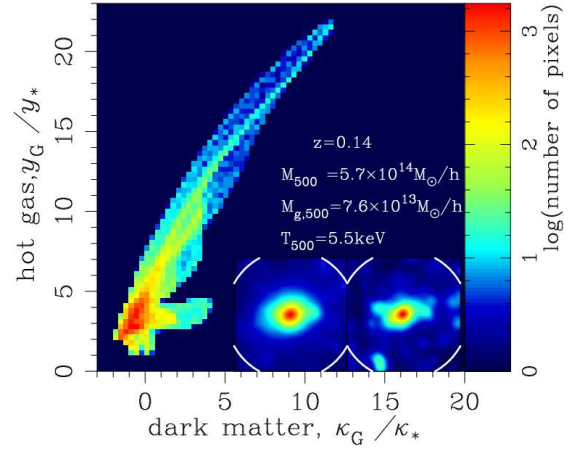}
\epsscale{1.}
\caption{Same as Fig.~1 but for a $25'\times 25'$ region 
of a single massive halo at $z=0.142$
found in the simulation. 
The halo properties such as $M_{500}$ and $T_{500}$
are also indicated within the figure.
Also plotted are
the $y_{\rm G}$ and $\kappa_{\rm G}$ maps smoothed with $\theta_{\rm
 fwhm}=2'$ Gaussian in the left and the right subpanels, respectively.
A circle with radius of $\theta_{500}$ is shown in each subpanel.
}
\label{fig:bfd_halo}
\end{figure}


\begin{figure}
\epsscale{0.85}
\plotone{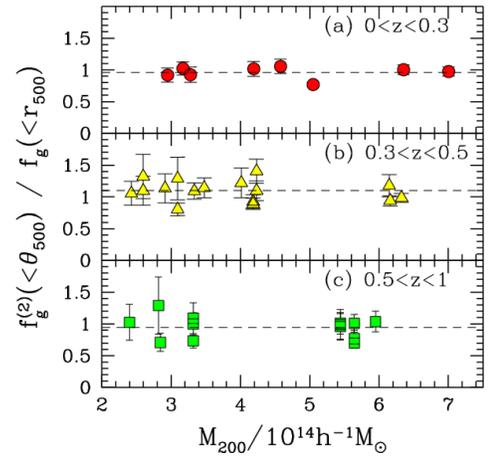}
\caption{Distributions of the ratio 
$f_g^{(2)}(<\theta_{500})/f_g(<r_{500})$
between the projected (Eq.~[5]) and true gas fractions against the halo mass
 $M_{200}$ for three redshift subsamples,
obtained with the same halo selection as in Fig.~1
but from a total survey area of $\approx 20{\rm deg}^2$.
Each data point is measured from a pair of the Gaussian-smoothed 
{\it noise-free} $\kappa_{\rm G}$ and $y_{\rm G}$ maps,
 while each error-bar represents an observational uncertainty ($1\sigma$)
based on $50$ Monte-Carlo realizations of random noise in $\kappa_{\rm
 G}$ and $y_{\rm G}$.
The sample means
over the noise-free measurements are indicated as dashed lines.}
\label{fig:kap}
\end{figure} 

The butterfly diagram of a single cluster shows a wing
   that contains rich information on the relative distributions
   of dark matter and hot gas.
The tip of a butterfly wing shows the surface mass densities of both
dark matter and hot gas in the cluster core, and the bulk of the wing
reveals the degree to which the hot gas traces dark matter.
The wing width serves as an indicator of the difference 
between projected distributions of dark matter and hot gas.
Most halos shown in Fig.~1 have significant wing widths. These
   widths arise because the baryon distribution is closer to 
   axisymmetric than the dark matter distribution due to collisional
   relaxation of the baryons.
The local slope  $\eta_g(\btheta)$ of a wing
in conjunction with the cluster redshift and temperature
can be
an indicator of the local gas fraction $f_g(\btheta)$ (see Eq.~[\ref{eq:y2k}]).
Note that the estimator $\eta_{g}^{(2)}$ 
defined in Eq.~(\ref{eq:cor}) can be interpreted as  a global slope 
of a wing in the least-square sense. 

We use the halo sample with $\kappa_{\rm G}\simgt 4.5\kappa_*$ and
 $y_{\rm G}\simgt 6y_*$ 
derived from our noise-free sky maps of $29{\rm deg}^2$
to compare 
the halo-based cumulative gas fractions
$f_g(<r)=M_g(r)/M(r)$ 
with
their projected estimates
$f_g^{(2)}(<\theta)\propto \eta_g^{(2)}(<\theta) T_g^{-1}
\Sigma_{\rm crit}^{-1}(z_d,z_s)$ from cross correlation
between $\kappa$ and $y$;
$\theta=r/D_A(z_d)$ 
with $D_A(z)$ being the angular diameter distance 
to redshift $z$.
We make this comparison inside $r_{500}$, and calculate for each
   halo a gas fraction using Eqs.~(\ref{eq:eta}) and (\ref{eq:cor})
   within an aperture of projected radius
   $\theta_{500}=r_{500}/D_A(z_d)$. We assume cluster
   isothermality and take the emission-weighted mean temperature
   $T_{500}$ inside $r_{500}$ as the observable X-ray temperature.
The weight $w_i$ in Eq.~(\ref{eq:eta}) is set to zero if
$\kappa_i\le 0$ or $y_i\le 0$ to exclude low-signal regions
where cross correlation measurements are of low significance due to
projection effects and/or observational noise.
Further, we remove those halos 
whose centers are located outside the inner $50'\times 50'$ region 
of a $1{\rm deg}^2$ sky map, in order to avoid systematics in the gas
fraction estimates due to 
the boundaries. The effective survey area is thus about
   $20\ {\rm deg}^2$

Figure~3 shows the distribution of the ratio $ f_g^{(2)}
   (<\theta_{500})/ f_g (<r_{500})$ obtained from the noise-free maps
   as a function of $M_{200}$ for three redshift subsamples:
   (a) $z\le 0.3$ (where we find 8 halos); (b) $0.3\le z<0.5$ (16
   halos); and (c) $0.5\le z<1$ (13 halos).
The sample means and their standard errors
of $f_g^{(2)}(<\theta_{500})/f_g(<r_{500})$ 
are (a) $0.96\pm 0.03$, (b) $1.10\pm 0.04$, and (c) $0.95\pm 0.05$.
Note that the quoted errors represent
the intrinsic scatter 
of the estimator $f_g^{(2)}$
from the noise-free signal maps.
On the whole, the estimator $f_g^{(2)}(<\theta_{500})$ for the gas
   fraction is close to unbiased, but there is a slight
   ($\sim 2\sigma$) over-estimation bias of $f_g$ in 
   $ 0.3 \le z < 0.5$. 
We found that most halos with overestimated values of $f_g$ are
   associated with mergers. The shock heating of gas in mergers boosts
   the temperature in the region between the merging halos. However,
   we ascribe the lower temperature $T_{500}$ to this gas, based on the
   emission-weighted temperature of the entire halo, and so
   overestimate the value of $f_g$ for the system as a whole.
   This overestimate of $f_g$ appears to be smaller at higher $z$
   because such halos have smaller angular sizes and so suffer larger
   beam smearing effects. If the 5 interacting/merger systems in
   subsample~(b) are rejected, then the sample mean
   $f_g^{(2)}(<\theta_{500})/f_g(<r_{500}) = 1.01\pm 0.03$, showing
   that the bias is eliminated. Such a debias should be possible based
   on X-ray observations of the survey field.
%
%

   On the other hand, the projection effect in $\kappa$ due to
   intervening matter (Metzler et al. 2001) also affects cluster 
   $f_g$ estimates made using weak lensing data. This is more severe
   for higher-$z$ clusters due to their lower geometrical lensing
   efficiency and smaller angular sizes. If an intervening mass
   concentration is projected onto a cluster, then the gas fraction 
   of that cluster could be under-estimated. Furthermore, the
   gradients in 3D cluster gas-fraction and temperature profiles
   (Ettori et al. 2004) can also induce a systematic bias 
   in projected $f_g^{(2)}$ estimates, especially for
   spatially-resolved low-$z$ clusters. Intrinsic gradients in $T_g$
   and $f_g$ are manifest as a curvature of the wing in a butterfly
   diagram, and they cannot be distinguished unless detailed
   information is available on the temperature distribution in the ICM.

The error bars in Fig.~(3) are based on 50 realizations of 
   Monte-Carlo simulated noise maps which are linearly added to
   the Gaussian smoothed signal maps of $\kappa_{\rm G}$ and 
   $y_{\rm G}$. Following van Waerbeke (2000), a correlated noise map
   of $\kappa_{\rm G}$ is generated for each realization
   using the noise auto-correlation function specified 
   by the smoothing scale $\theta_{\rm fwhm}$ and the observational
   noise rms $\sigma_{\kappa}$ (Eq.~[\ref{eq:knoise}]). A similar
   correlated noise map for $y_{\rm G}$ uses the noise rms
   $\sigma_{y}$ (Eq.~[\ref{eq:ynoise}]). For a fixed halo mass, 
   the random error in $f_g^{(2)}$ is smaller for lower-$z$ halos with
   larger angular scales, $\theta_{500}$, over which pixel noise is
   averaged, but systematic errors will be dominant for these clusters.


\section{Conclusions}\label{sect:conclusion}

By simulating the AMiBA experiment, we have shown that the
   butterfly diagram that emerges from a comparison of the SZ and
   weak-lensing signals contains rich information on the
   distribution of the cluster gas fraction within an individual
   cluster and between different clusters (\S 4). 
   The cross-correlation based estimator $f_g^{(2)}(<\theta)$ should
   be used for a robust estimate of the cluster gas fraction. 
   Overall, this estimator is close to unbiased, but some halos show
   intrinsic scatter that exceeds the observational errors (\S 3)
   due to various sources of systematics (\S 4). The bias for the 
   the intermediate-$z$ subsample of halos can be eliminated by
   removing interacting/merger systems. Careful selection of clusters
   should minimize systematic uncertainties in $f_g$. Detailed
   information regarding the temperature distribution in clusters, 
   derived from X-ray observations, improves the calibration
   of the gas fraction estimator, and yields valuable information on
   the thermodynamic history of the ICM.


\acknowledgments 
We acknowledge Kai-Yang Lin and Masahiro Takada for useful discussions.


\begin{thebibliography}{}

\bibitem[]{}

\bibitem[]{BS01}
Bartelmann, M., \& Schneider, P., 2001, Phys.~Rept, 340, 291	

\bibitem[]{}
Bialek, J.J., Evrard, A.E., \& Mohr, J.J. 2001, 555, 597

\bibitem[]{}
Birkinshaw, M., 1999, Phys. Rep., 310, 97

\bibitem[]{}
Broadhurst, T., Taylor, A.N., \& Peacock, J.A. 1995, ApJ, 438, 49

\bibitem[]{}
Broadhurst, T., Takada, M., Umetsu, K., Kong, X., Arimoto, N., Chiba,
	  M., \& Futamase, T. 2005, ApJ, 619, L143







\bibitem[]{}
Ettori, S., Tozzi, P., Borgani, S., \&  Rosati, P. 2004, A\&A, 417, 13

\bibitem[]{}
Hamana, T., Takada, M., \& Yoshida, N., 2004, MNRAS, 350, 893

\bibitem[]{Ho00}
Ho, P.T.P. et al., 2004, MPLA, 19, 13-16, 993


\bibitem[]{}
Holder, G.P., Carlstrom, J.E., \& Evrard, A.E. 2000,
	{\it Constructing the Universe with Clusters of Galaxies}, 
	  IAP 2000 meeting, Paris, France

\bibitem[]{}
Ishizaka, C. \& Mineshige, S. 1996, PASJ, 48, L37


\bibitem[]{}
Kravtsov, A.V., Nagai, D., \& Vikhlinin, A.A. 2005, ApJ, 625, 588


\bibitem[]{Lin04}
Lin, K.-Y.,  Lin, L., Woo, T.-K., Tseng, Y.-H., \& Chiueh, T.,

\bibitem[]{}
Metzler, C.A., White, M., \& Loken, C. 2001, ApJ, 547, 560

\bibitem[]{}
Myers, S.T., Baker, J.E., Readhead, A.C.S., Leitch, E.M., \& Herbig, T.
1997, 485, 1

\bibitem[]{}
Miyazaki, S. et al., 2002, ApJ, 580, L97

\bibitem[]{}
Park, C.-G., Ng, K.-W., Park, C., Liu, G.-C., \& Umetsu, K. 2003,
 	ApJ,  589, 67

\bibitem[]{}
Pearson, T.J. et al. 2003, ApJ, 591, 556


\bibitem[]{}
Piffaretti, R., Jetzer, P., Kaastra, J.S., \& Tamura, T. 2005,  
A\&A, 433, 101



\bibitem[]{} Schneider, P. \& Seitz, C. 1995, 294, 411

\bibitem[]{GADGET}
Springel, V., Yoshida, N., \& White, S.~D.~M. 2001, NewA, 6, 79

\bibitem[]{Umetsu04}
Umetsu, K., Chiueh, T., Lin, K.-Y., Wu, J.-M., \& Tseng, Y.-H.,
2004, MPLA, 19, 13-16, 1027

\bibitem[]{}
van Waerbeke, L 2000, MNRAS, 313, 524


\bibitem[]{}
White, S.D.M. et al. 1993, Nature, 366, 429

\bibitem[]{}
Zhang, P., Pen, U.-L., \& Benjamin, W. 2002, ApJ, 577, 555


\end{thebibliography}
\end{document}